\documentclass[aps,prb,twocolumn,showpacs,preprintnumbers,superscriptaddress]{revtex4}
\usepackage{graphicx}
\usepackage{dcolumn}
\begin{document}

\title{Transverse and Longitudinal Magnetic-Field Responses in the Ising Ferromagnets URhGe, UCoGe and UGe$_{2}$}

\author{F. Hardy}
\email[]{Frederic.Hardy@kit.edu}
\affiliation{Karlsruher Institut f\"ur Technologie, Institut f\"ur Festk\"orperphysik, 76021 Karlsruhe, Germany}
\author{D. Aoki}
\affiliation{Commissariat \`a l'\'Energie Atomique, INAC/SPSMS, 38054 Grenoble, France}
\author{C. Meingast}
\author{P. Schweiss}
\author{P. Burger}
\affiliation{Karlsruher Institut f\"ur Technologie, Institut f\"ur Festk\"orperphysik, 76021 Karlsruhe, Germany}
\author{H. v. L\"ohneysen}
\affiliation{Karlsruher Institut f\"ur Technologie, Institut f\"ur Festk\"orperphysik, 76021 Karlsruhe, Germany}
\affiliation{Karlsruher Institut f\"ur Technologie, Physikalisches Institut, 76128 Karlsruhe, Germany}
\author{J. Flouquet}
\affiliation{Commissariat \`a l'\'Energie Atomique, INAC/SPSMS, 38054 Grenoble, France}

\date{\today}
\begin{abstract}
The field dependence of the electronic specific heat $\gamma$(H) of URhGe is determined using temperature-dependent magnetization measurements and Maxwell's relation for all three orthorhombic directions. A large ($\approx$ 40 \%) enhancement of $\gamma$(H) is found at the reorientational transition for fields applied along the {\it b} hard axis, which we argue occurs when the field-induced moment is of the same size as the zero-field ordered moment M$_{0}$. Fields applied along the easy direction, on the other hand, lead to a rapid suppression of $\gamma$(H). We emphasize that this behavior is also applicable to the closely related ferromagnetic superconductors UCoGe and UGe$_{2}$ and discuss implications for the novel field-induced/enhanced superconductivity in these materials.
\end{abstract}
\maketitle


%
Due to the weakness of the characteristic energies, magnetic fields (H) can induce drastic changes in the ground-state properties of heavy-fermion systems, particularly on the magnetic fluctuations, which play a crucial role in the dressing of the quasiparticle effective mass $m^{*}$. For instance, in systems dominated by antiferromagnetic correlations, magnetic-field tuning leads to quantum criticality, pseudo-metamagnetism or paramagnetic depairing and can reveal a new interplay between antiferromagnetism (AFM) and superconductivity (SC).~\cite{Flouquet05,HvL07, Flouquet10} Recently, spectacular effects, {\it e.g.} field-induced or enhanced SC,~\cite{Levy05,Aoki09} have been reported in the Ising-like ferromagnets (FM) URhGe and UCoGe, indicating that magnetic field can also cause strong modifications of their ferromagnetic properties. In particular, recent resistivity studies~\cite{Miyake08} suggest that the field-induced SC can be explained by a field-induced enhancement of the effective mass, which has, however, not been determined directly. 

In this Letter, we report detailed magnetization measurements on a URhGe single crystal with accurate field alignment, from which we can derive directly the change of the effective mass for both transverse and longitudinal magnetic fields. In particular, we determine (i) the magnetization curves M(0,H) at T=0 and (ii) the field dependence of both the Curie temperature T$_{\rm Curie}$(H) and the Sommerfeld coefficient $\gamma (H)$ using the Maxwell relation $\left(\partial M/\partial T\right)_{H}$=$\left(\partial S/\partial H\right)_{T}$. For comparison with UCoGe which has a lower $T_{\rm Curie}$ ($\approx 2.7\,{\rm K}$) than URhGe (9.5 K), direct specific-heat measurements were performed down to $0.4\,{\rm K}$ for both compounds. The present data are compared with previous resistivity measurements used to determine the field dependence of the $A(H)$ coefficient.~\cite{Miyake08,Aoki09}
\begin{figure}[b]
\begin{center}
\includegraphics[width=8.5cm]{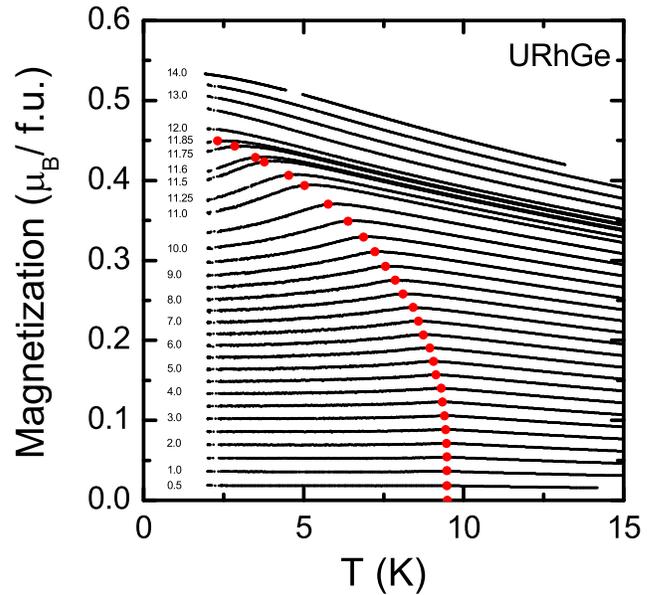}
\caption{\label{fig:Fig1} (Color online) Magnetization as a function of temperature for URhGe in different fields applied parallel to the crystal {\it b}-hard axis. The continuous transition line, T$_{\rm Curie}$(H), appears as a maximum in the magnetization (red symbols).}
\end{center}
\end{figure}
Our results indicate that an important parameter is the strength of the field-induced transverse moment with respect to the ordered moment M$_{0}$. For H $\|$ M$_{0}$, magnetic fluctuations are progressively suppressed by the field and the strength of the SC pairing is weakened, while for H $\perp$ M$_{0}$, magnetic field causes a decrease of the Curie temperature and an enhancement of both FM fluctuations and SC pairing strength.

High-quality single crystals of URhGe were grown by the Czochralski method using a tetra-arc furnace.~\cite{Aoki11_ICHE} Magnetization measurements were performed with a Vibrating Sample Magnetometer (VSM) down to $2\,{\rm K}$ in magnetic fields up to $14\,{\rm T}$.

\begin{figure}[h]
\begin{center}
\includegraphics[width=7.5cm]{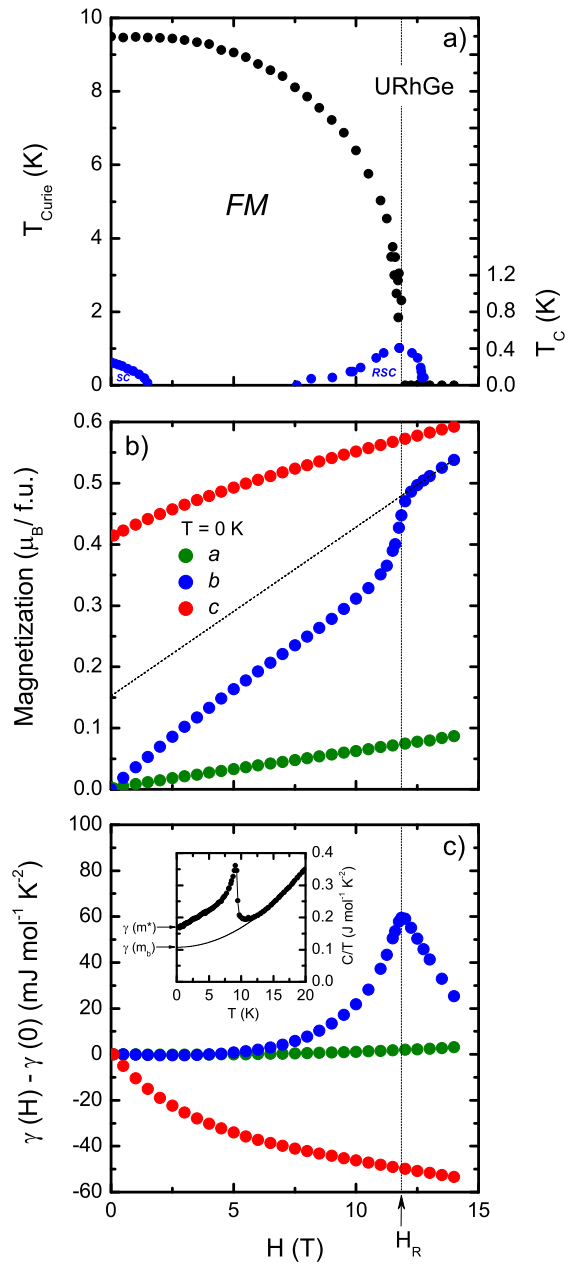}
\caption{\label{fig:Fig2} (Color online) a) Field dependence of T$_{\rm Curie}$ (black symbols, left-hand scale) and T$_{c}$ (blue symbols and right-hand scale, taken from Ref.~\onlinecite{Miyake08}) for H $\|$ {\it b} obtained from magnetization and transport measurements, respectively. b) Zero-temperature magnetization curves for H $\|$ {\it a}, {\it b} and {\it c}. c) Field dependence of $\gamma^{*}$(H) derived using the Maxwell relation for H $\|$ {\it a}, {\it b} and {\it c}. The inset shows the specific heat for H=0. The solid line is a fit to $\gamma(m_{b})$+B$_{3}$T$^{2}$ for T $>$ T$_{\rm Curie}$.}  
\end{center}
\end{figure}

Figure \ref{fig:Fig1} shows the magnetization M(T,H) of URhGe as a function of temperature in several constant fields applied along the hard magnetization axis {\it b}, the easy direction being the {\it c} axis. The ferromagnetic transition is easily detected in our measurements. Indeed, when the transition line is crossed, the magnetization exhibits a maximum that becomes more pronounced with increasing fields. As reported previously,~\cite{Levy05,Miyake08,Lev07} the transition is continuous for H $<$ 12 T, and the temperature T$_{\rm Curie}$(H) where it occurs decreases smoothly with increasing field parallel to {\it b}. Moreover, when the component induced parallel to {\it b} is small, the decrease of T$_{\rm Curie}$ is quadratic with field, as proved theoretically in the mean-field limit.~\cite{Min10,Min10_Arx} The relatively high value of $T_{\rm Curie}$ allows to determine the 0 K magnetization curve M(0,H) by fitting the low-temperature part of the M(T,H) data with the Fermi-liquid expression,
\begin{equation}\label{eq:eq2}
M(T) = M_{0}-\beta T^{2},
\end{equation}
At $2\,{\rm K}$, the $T_{\rm Curie}/T$ ratio $\sim 4$ is large and allows to observe the $T^2$ dependence of $M(H)$. Figure \ref{fig:Fig2}b shows the extrapolated values M(0,H) obtained for H $\|$ {\it b} as well as for H $\|$ {\it a} and {\it c}. The application of a field perpendicular to the easy {\it c} axis, {\it i.e.} in the {\it b} direction, at zero temperature destabilizes the FM state, as demonstrated in Fig.\ref{fig:Fig2}a. Fig.\ref{fig:Fig2}b shows that the magnetization along {\it b} increases strongly with field due to the quite large initial susceptibility $\chi_{b}$ and then increases sharply upon crossing the phase boundary at about 12 T from the FM phase to the polarized phase. This is the field at which the field-induced SC is most robust. We note that this phase transition occurs when the field-induced moment M(H) is approximately equal to the zero-field ordered moment along the easy direction, which will be discussed in more detail later. Interestingly, a linear extrapolation of M(0,H) from H $>$ H$_{R}$ to H = 0 exhibits a non-zero intercept, suggesting that the high-field phase also possesses a finite ordered moment. This illustrates that the application of H perpendicular to the easy axis moves the system towards a FM instability. This tendency is clearly illustrated by the field dependence of T$_{\rm Curie}$ shown in Fig.\ref{fig:Fig2}a for H ${\|}$ {\it b}. Here the accurate orientation of the magnetic field with the {\it b} axis is confirmed by the observed low value of H$_{R}$ which is extremely sensitive to field misalignment.~\cite{Aoki11_ICHE}
\begin{table}[h]
\caption{\label{tab:Table1} Properties of the three ferromagnets URhGe, UCoGe and UGe$_{2}$ at ambient pressure. $p_{c}$ is the critical pressure where FM disappears.}
\begin{ruledtabular}
\begin{tabular}{c|ccccc}
                & easy                    & $T_{\rm Curie}$           &  $M_{0}$               & $\gamma$              &  $p_{c}$       \\
                & axis                    & (K)                 & ($\mu_{B}$)             & (${\rm mJ\,mol^{-1}K^{-2}}$) &  (GPa)       \\
\hline    
{\bf URhGe}		  & {\it c}							    & 9.5						        & 0.4		        				 &  163                      &  $>$13         \\

{\bf UCoGe}		  & {\it c}							    &	2.7					          & 0.07                   &  57                       &  1.5           \\

{\bf UGe$_{2}$}	&	{\it a}							    &	52					          & 1.48                   &  34                       &  1.6                 
\end{tabular}
\end{ruledtabular}
\end{table}

\begin{figure}[h]
\begin{center}
\includegraphics[width=8.5cm]{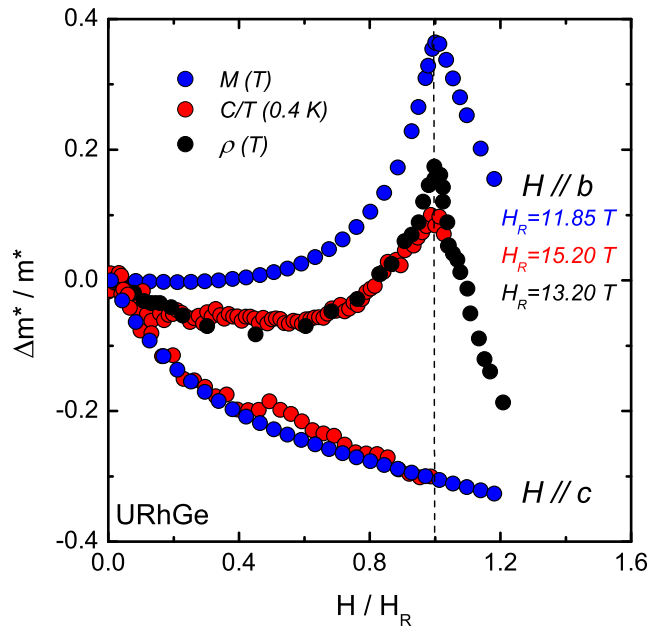}
\caption{\label{fig:Fig3} (Color online) Comparison of the field dependence of the effective mass $m^{*}$ for H $\|$ {\it b} and H $\|$ {\it c} obtained from magnetization, specific-heat and transport measurements.}
\end{center}
\end{figure}

Figure \ref{fig:Fig2}c shows the field dependence of the specific-heat linear coefficient $\gamma(H)$, which is proportional to the average effective mass $m^{*} (H)$. It is calculated, for the three axes, by using the following Maxwell relation in the T $\rightarrow$ 0 limit, \begin{equation}\label{eq:eq3}
\left(\frac{\partial \gamma}{\partial H}\right)_{T} =  \left(\frac{\partial^{2} M}{\partial T^{2}} \right)_{H}= -2\beta,      
\end{equation}
where $\beta$ is taken from the fit of the M(T,H) curves to Eq. \ref{eq:eq2}. The resulting data are compared in Fig. \ref{fig:Fig3} to the dependence of $m^{*}$(H) obtained from transport~\cite{Miyake08} and direct specific-heat measurements. For H $\|$ {\it b}, the strong increase of $\gamma(H)$ at H$_{R}$ is in good agreement with a previous derivation of $m^{*}$(H) obtained from the coefficient of the temperature dependent part of the resistivity A(H) $\propto$ $\left(m^{*}(H)\right)^{2}$ (see Fig.\ref{fig:Fig3}). On the other hand, for H$\|${\it c}, a continuous decrease of $\gamma (H)$ related to the suppression of longitudinal fluctuations is observed, while only a tiny increase is observed for H ${\|}$ a. The relatively good agreement between the different data for H $\|$ {\it b} shows that (i) our procedure to derive $\gamma (H)$ is correct since, for H $\|$ {\it c}, it accurately reproduces direct specific-heat measurements, and (ii) a rigorous alignment of H with the crystal {\it b}-axis is crucial to strongly enhance $m^{*}$ and, as discussed in Refs.~\onlinecite{Aoki11_ICHE} and \onlinecite{Lev07}, to promote reentrant SC with field. Indeed, small misalignements with respect to the {\it b} axis shift H$_{R}$ to higher fields and lower the enhancement of $m^{*}$. 

Following Miyake {\it et al.},~\cite{Miyake08} the effective mass $m^{*}$(H) can be described by
\begin{equation}\label{eq:eq4}
m^{*}(H)=m_{\rm b}+m^{**}(H),\\
\end{equation}
where $m_{\rm b}$ is the renormalized band mass and $m^{**}$(H) is the FM correlated mass associated with the FM instability. 
For URhGe, we can estimate these quantities in zero field using specific-heat measurements 
(see inset of Fig.\ref{fig:Fig2}). 
From the $T \rightarrow 0$ limit, we find that $\gamma (0) \approx 163 \,{\rm mJ\,mol^{-1} K^{-2}}$ 
while $\gamma (m_{\rm b}) \approx 110 \,{\rm mJ\,mol^{-1}K^{-2}}$ is obtained from a fit to $\gamma(m_{b}) + B_3 T^2$ (with $B_3=0.6\,{\rm mJ\,mol^{-1} K^{-4}}$) for $T > T_{\rm Curie}$. 
Thus, a large value of $\gamma(0)$ does not necessarily signal the proximity to a ferromagnetic instability since the absolute value of $\gamma$ depends also on the value of $\gamma(m_{\rm b})$ which can be rather high. It is the mass enhancement caused by the ferromagnetic correlations, $\gamma_{0}(m^{**})$ $\approx$ 53 mJ mol$^{-1}$ K$^{-2}$,  that really matters. Here, we have implicitly assumed that $\gamma(m_{b})$ is field independent. As shown in Fig.\ref{fig:Fig2}c, this appears to be a reasonable approximation for URhGe since $\gamma^{*}$(H) recovers the sole band-mass value $\approx$ $\gamma(m_{b})$ in the limit of very high fields ($H>16\,{\rm T}$) for $H\parallel c$.

\begin{figure}[tbh]
\begin{center}
\includegraphics[width=0.8 \hsize]{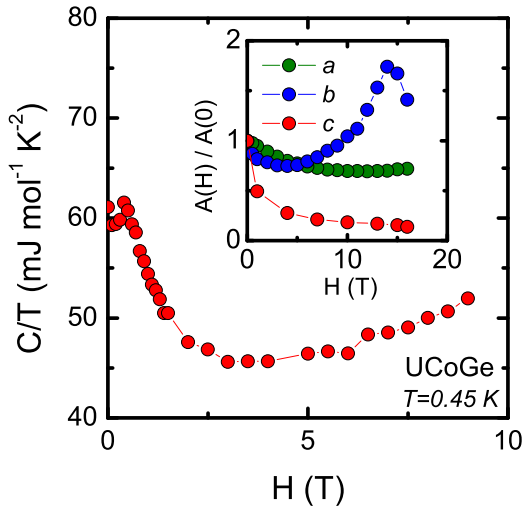}
\caption{\label{fig:UCoGe} (Color online) Field dependence of $C/T$ for $H\parallel c$ at $0.45\,{\rm K}$ in UCoGe.
The inset shows relative field variation of $A(H)/A(0)$ for the field along $a$, $b$ and $c$-axis, cited from Ref.~\onlinecite{Aoki09}}
\end{center}
\end{figure}

In the following, we argue that this field-induced enhancement of $\gamma$ is quite general and should also occur in UCoGe and UGe$_{2}$. We expect a maximum of $m^{*}$ at the field where the moment along the hard direction is of the order of the ordered moment at zero field, and the low-field susceptibilities can be used to quantify this field for the different compounds. That is, the enhancement of $m^{*}$ will occur at H$_{b}$=M$_{0}$/$\chi_{b}$ and H$_{a}$=M$_{0}$/$\chi_{a}$.   
To test these predictions, we have calculated the characteristic fields H$_{i}$ ({\it i}={\it a}, {\it b} and {\it c}) for the three ferromagnetic superconductors URhGe, UCoGe and  UGe$_{2}$. These values, together with susceptibilities $\chi_{i}$, are summarized in Table \ref{tab:Table2}. The peak of $m^{*}$(H) at 12 T in URhGe and the maximum of A(H) around 15 T in UCoGe (see inset of Fig.~\ref{fig:UCoGe}) coincide quite well with the derived values of H$_{b}$. Moreover, the observed rapid quenching of the ferromagnetic correlations for H$_{c}$ $\approx$ 2.5 T in UCoGe can be related to the abrupt changes in the field variation of $C/T$ and of the $A$ coefficient~\cite{Aoki09} measured for $H\parallel c$ (see Fig.~\ref{fig:UCoGe}). 

Furthermore, the large value of M$_{0}$ in UGe$_{2}$, explains why no drastic effects were observed in the transverse and longitudinal responses up to 30 T.~\cite{Sakon07} However, the magnetic field parallel to the {\it a} easy axis can affect the SC pairing near p$_{x}$ $\approx$ 1.2 GPa, where the system switches from the low-moment FM1 to the high-moment FM2 phases and where T$_{c}$($p$) and $\gamma$(p) are maximum.~\cite{Pfleiderer02,Taufour10,Tateiwa01,Huxley01} For $p$ $>$ $p_{c}$, the situation is completely different 
since tricriticality is accompanied with a field reentrance in the FM1 phase through a first-order metamagnetic transition which will end up at a quantum critical end-point (QCEP), $p_{\rm QCEP}\sim 2 p_{\rm c}$ and $H_{\rm c}\sim 16\,{\rm T}$.~\cite{Taufour10,Kotegawa11} 
As UCoGe becomes FM already at $p=0$ through a first order transition associated with a possible mixing of FM and PM phases depending on the crystal growth conditions~\cite{Oht08}, similar effects can be expected. They can be the origin of the small sharp maxima of $C/T$ for $H\parallel c$ (see Fig.~\ref{fig:UCoGe}). However, the weakness of $M_0$ may explain why no large initial field-enhancement of $C/T$ occurs.
Basically, in UCoGe, $p_{\rm QCEP}$ may be close to $p_{\rm c}$ and $H_{\rm QCEP}$ may be low.
For URhGe, tricriticality does not need to be considered as the FM transition is of second order 
and pressure drives the system deep inside the FM region.~\cite{Hardy05c}

\begin{table}[h]
\caption{\label{tab:Table2} Susceptibilities and characteristic fields of the three ferromagnetic superconductors URhGe, UCoGe and UGe$_{2}$. Data for UGe$_{2}$ and UCoGe were taken from Refs.~\onlinecite{Sakon07,Huy08}, respectively.} 
\begin{ruledtabular}
\begin{tabular}{c|ccc|ccc}
                & $\chi_{a}$              & $\chi_{b}$              &  $\chi_{c}$              & $H_{a}$                   &  $H_{b}$         &  $H_{c}$\\
                
                & \multicolumn{3}{c|}{($\mu_{B}$ $T^{-1}$)}            & \multicolumn{3}{c}{($T$)}\\
\hline    
{\bf URhGe}		  & 0.006							      & 0.03						        & 0.01		        				 &  66                       &  13              &40\\

{\bf UCoGe}		  & 0.0024						      &	0.006					          & 0.029                    &  29                       &  12              &2.5\\

{\bf UGe$_{2}$}	&	0.006							      &	0.0055					        & 0.011                    &  230                      &  250             &122             
\end{tabular}
\end{ruledtabular}
\end{table}
In conclusion, in URhGe and UCoGe, we have shown that an applied transverse field drives the system through a FM instability leading to a strong enhancement of the correlated effective mass. We find roughly a 40 \% enhancement of $\gamma$(H) at the field-induced FM instability, which must be related to a proliferation of magnetic fluctuations. The detailed nature of these fluctuations will have to be investigated using neutron scattering in the future, however our results suggest, as pointed out in Ref.~\onlinecite{Levy09}, that the spins fluctuate mostly in the (b,c) plane, since the transition occurs when the magnetization in the {\it b} direction is equal to M$_{0}$. In a sense, the magnetic field compensates the magnetic anisotropy opening the way to these fluctuations. The reported longitudinal and transverse field variations of $m^{*}$ have also a strong feedback on the field dependence of $H_{\rm c2}$. Theoretical approaches have been derived recently in weak and strong coupling models.~\cite{Min10,Tad10}
In the first case, it is stressed that, in a two-band model inside the FM boundary, the magnetic field applied along the easy axis can lead to the collapse of the assumed dominant $\uparrow\uparrow$ spin pairing. In the strong-coupling limit, the unusual enhancement of $H_{\rm c2}$ is a combined effect of the field dependence of the FM fluctuations and of the position of the nodes. Now new goals will be to observe the phenomena on the spin dynamics and to analyze carefully the contrast between Heisenberg and Ising itinerant ferromagnets. For the Heisenberg ferromagnet ZrZn$_2$, the field dependence of the NMR spin-lattice relaxation time indicated that the magnetic field wipes out the FM fluctuations.~\cite{Kon76}  The novelty of the Ising itinerant ferromagnets is the transverse response with the field enhancement of $m^*$. 

We thank V. Mineev for his theoretical input, and K. Asayama and H. Kotegawa for discussion on NMR studies. This work was supported by the Deutsche Forschungsgemeinschaft (under Grant No. FOR 960), the ERC starting grant (NewHeavyFermion) and the French ANR project (CORMAT, SINUS and DELICE).
\bibliography{biblio}
\end{document}